\documentstyle[aps,epsfig,prl,floats]{revtex}

\begin{document}
\draft
\title{Field dependent thermodynamics and Quantum Critical
Phenomena in  the dimerized 
 spin system Cu$_2$(C$_5$H$_{12}$N$_2$)$_2$Cl$_4$  }

\author{Norbert Elstner}
\address{Physikalisches Institut, Universit\"at Bonn, 
         Nu\ss allee 12, D-53115 Bonn, Germany }
\author{Rajiv R. P. Singh}
\address{Department of Physics, University of California, 
Davis, California 95616}

\twocolumn[
\date{\today}
\maketitle
\widetext
\begin{abstract}
  \begin {center}
    \parbox{6in}{ 
      Experimental data for the uniform susceptibility, magnetization
      and specific heat  for the material
      Cu$_2$(C$_5$H$_{12}$N$_2$)$_2$Cl$_4$ (abbreviated CuHpCl)
      as a function of temperature and external field
      are compared with those of three different dimerized spin
      models: alternating spin-chains, spin-ladders and the bilayer
      Heisenberg model. It is shown that because this material consists
      of weakly coupled spin-dimers, much of the data is 
      insensitive to how the dimers are coupled together and what the
      effective dimensionality of the system is. When
      such a system is tuned  to the quantum critical point by application
      of a field, the dimensionality shows up 
      in the power-law dependences of thermodynamic quantities on temperature. 
      We discuss the temperature window for such a quantum critical
      behavior in CuHpCl.}
  \end{center}
\end{abstract}

\pacs{\hspace{0.5in}PACS: 75.10.Jm, 75.40.Cx, 75.40.Mg, 75.50.-y, 75.50.Ee}

]

\narrowtext

\section{Introduction} 

Quantum disorder and spin-gap phenomena in insulating and doped
magnetic systems have attracted much interest recently. A number
of novel materials have been synthesized, which have a gap
in the spin excitation spectrum and consequently exhibit
activated thermodynamic behavior at low temperatures.
Examples of such systems include
CaV$_n$O$_{2n+1}$ \cite{taniguchi}, 
(VO$_2)$P$_2$O$_7$ \cite{VPO}, 
SrCu$_2$O$_3$ \cite{Azuma94}, 
CuGeO$_3$ \cite{CuGeO3}, 
Na$_2$Ti$_2$Sb$_2$O\cite{Kauzlarich}
and a number of organic materials such as
Cu$_2$(C$_5$H$_{12}$N$_2$)$_2$Cl$_4$ 
abbreviated as CuHpCl \cite{CuHpCl1,CuHpCl2}.
While the relationship of spin-gap behavior to high temperature
superconductivity remains one of the most intriguing problems in
condensed matter physics \cite{MM,Anderson}, the insulating
materials are of interest in their own right providing rich interplay
of quantum-chemistry, strong quantum fluctuations,
frustration and dimensionality crossovers.

The material CuHpCl is particularly interesting for a variety of reasons.
First, from the point of view of spin-half models, it is truely in the strong
coupling limit, where pairs of spins are strongly coupled to each other
forming a spin-dimer and these pairs are then weakly coupled to the rest of
the system. This provides a testing ground for strong coupling theories of the
quantum disordered phase. Secondly, the exchange energy scale is small enough
so that by application of an external magnetic field one can drive the system
through a phase transition from the quantum disordered to a magnetically
ordered phase. Third, there is potentially, a hierarchy of energy scales so
that there is a temperature window of low-dimensional quantum critical
behavior, which at the lowest temperatures will crossover to 3-dimensional
behavior.

We present here results of finite temperature strong coupling
expansions around dimerized Hamiltonians. We have developed a method
combining conventional many-body perturbation theory 
interaction representation with cluster expansion techniques to carry
out these expansions by fully automated computer programs.  Technical
details of these calculations will be presented elsewhere \cite{ES}.
At high temperatures, this approach is related to conventional high
temperature expansions. But, unlike those, the strong coupling
expansions show excellent convergence down to very low temperatures
for a range of parameters, and thus allow to reliably compare the
experimental data with model Hamiltonians over the entire temperature
range. We have calculated the uniform susceptibility, magnetization,
internal energy and specific heat of 3 different classes of model
Hamiltonians by this method. These include two quasi-one dimensional
models, the alternating spin-chain and the two-leg spin-ladder, and
one quasi-two dimensional model, the spin-bilayer. These models have
been the focus of numerous theoretical investigations in recent years
\cite{dagotto,sandvik,weihong}, however, ours is the first comprehensive 
numerical study of the field and temperature dependent thermodynamics
of these models.

The material CuHpCl consists of quasi-1D polymeric
chains\cite{Chiari}, where two spin-half Copper atoms are relatively
close to each other and are physically quite far from other pairs of
atoms.  Based on the linear polymeric structure, quasi-1D models of
weakly coupled spin-dimers have been favored for this system. However,
the physical distance for pairs of atoms between different polymers
and those within a given polymer are comparable. Furthermore, one can
identify exchange pathways that lead to interaction between spins on
different chains as well as those in a given chain. Thus, in the
absence of any first principles calculation, the spin Hamiltonian for
this system and the effective dimensionality is not apriori
obvious. The majority of the experimental data on the material has
been interpreted in terms of one-dimensional dimerized spin models
\cite{CuHpCl1,CuHpCl2,CuHpCl3}, where the exchange coupling for the
dimers is about 13K, whereas the coupling between spins on neighboring
dimers is about 3K. However, there are two pieces of experimental data
which are in conflict with this picture.  First, on application of a
strong magnetic field when the spin-gap vanishes, a finite temperature
transition has been observed at temperatures of order 1K, whereas no
such finite temperature phase transition can exist in a strictly
one-dimensional system. Secondly, neutron diffraction spectra on
powder samples do not show any evidence of Van-Hove singularities
expected from the quasi-1D models \cite{CuHpCl2}.

In this study, we find that the uniform thermodynamic quantities for
these models in zero field are very insensitive to the way in which
the dimers are coupled together. In particular, the susceptibility
data for this material is well described by either an alternating
chain model or a spin-ladder model or even a bilayer model. This may
appear surprising as the latter model is not apparently consistent
with the structure of the material.  The fact that one can
nevertheless obtain a good quantitative description of the
suceptibility measurements is a strong argument that thermodynamic
properties are determined by local effects thus rendering them almost
insensitive to the dimensionality.  This is easily understood in
leading order of perturbation theory: in a strongly dimerized system
the Curie-Weiss constant $\Theta_w$, which is the important parameter
controlling the high temperature susceptibility, and the singlet
triplet excitation gap $\Delta$, which is the relevant energy scale
controlling the low temperature susceptibility, are altered from their
non-interacting dimer values in the same combination independent of
how the dimers are coupled together.  Still, the extent to which one
is able to fit the experimental data by adjusting the parameters in
the three models is surprizing.  The theoretical calculations of the
specific heat in the three cases are also close to each other,
although they deviate from what is found experimentally.

It should be noted that the asymptotic low temperature behavior
of thermodynamic quantities do indeed depend on the spin-dispersion
along different directions. The dimensionality enters directly into
the power-law prefactors multiplying the activated behavior \cite{troyer}.
This study shows that such prefactors are essentially impossible
to see in strongly gapped systems.

We argue that a robust way to determine the dimensionality of
the spin system is by tuning it to the quantum critical point by
application of a magnetic field. The spin-gap in zero field, together
with the $g$-factors, determine the critical field. As the spin-gap
disappears, the thermodynamic quantities develop power-law
dependences on temperature \cite{CSY}. Given that these
materials are ultimately 3-dimensional, an important question is:
Is there a temperature window where the low dimensional quantum
critical behavior can be observed. We find that
at temperatures above $1K$, where our expansions still show good
convergence, the asymptotic low temperature power-laws are difficult
to determine in an unbiassed manner. However, in a coarse sense,
quantum critical behavior begins to set in at temperature of
order $5K$. By studying consistency with expected behavior in
one and two dimensions, one can determine the dimensionality
of the systems below this temperature. This crossover temperature
scale for the onset of power-laws is consistent
with recent NMR measurements \cite{CuHpCl3}.

The quasi-1D models that we wish to study are given by the Hamiltonian:
\begin{eqnarray}
   \label{general1d}
   {\cal H} &=& J_\perp \sum_{i} {\bf S}_{A,i} \cdot {\bf S}_{B,i} \\
            &+& J_l \sum_{i} [{\bf S}_{A,i} \cdot {\bf S}_{A,i+1} 
                        +{\bf S}_{B,i} \cdot {\bf S}_{B,i+1}] \nonumber \\
            &+& J_a \sum_{i} {\bf S}_{B,i} \cdot {\bf S}_{A,i+1} \nonumber
\end{eqnarray}
Here, ${\bf S}_{A,i}$ and ${\bf S}_{B,i}$ represent the two spins in unit cell
$i$ of the chain. The antiferromagnetic intra dimer coupling $J_\perp > 0$ is
much stronger than the exchange constants $J_l$ and $J_a$ between different 
dimers. We will focus on the two limiting cases of this general
model: the alternating spin chain with ($J_l = 0$ , $J_a \ne 0$) and the spin
ladder defined by ($J_l \ne 0$ , $J_a = 0$). 

In order to compare with a quasi-2D system, we consider
the bilayer Heisenberg model, with Hamiltonian: 
\begin{equation}
   \label{bilayer}
   {\cal H} = J_\perp \sum_{i} {\bf S}_{A,i} \cdot {\bf S}_{B,i} 
            + J_2 \sum_{<i,j>} [{\bf S}_{A,i} \cdot {\bf S}_{A,j} 
                                +{\bf S}_{B,i} \cdot {\bf S}_{B,j}] \;\;, 
\end{equation}
where $<i,j>$ are nearest neighbours on a square lattice, and A and B
represent spins in the two layers.
The coupling of the system to an external field is described by
\begin{equation}
   \label{ext-field}
   {\cal H}_{ext} = -g\mu_B H \sum_i \left(S^z_{A,i} + S^z_{B,i} \right)
\end{equation}

We perform numerical investigations of these models by applying finite
temperature strong coupling expansions \cite{ES}. 
The expansion parameter $\lambda$ is
given by ratio of the inter dimer to the intra dimer coupling, 
i.~e. $\lambda = J_a/J_\perp \, , \; J_l/J_\perp \, , \; J_2/J_\perp$
for the alternating chain, ladder and bilayer model respectively. 
The coefficients of the expansion are polynomials in the variables 
$J_\perp/k_{\rm B}T$, 
$1/f = 1/\left(1+\exp[-g\mu_BH/k_{\rm B}T]+\exp[g\mu_BH/k_{\rm B}T]\;\right)$
and $1/Z_0 = 1/\left(1 + f \exp[-J_\perp/k_{\rm B}T]\;\right) \;$, 
where $Z_0$ is the partition function of an isolated dimer.

Series for the uniform susceptibility $\chi$, the Magnetization $M$,
the internal energy $E$ and the specific heat $C$ 
are calculated complete to order $\lambda^8$, for arbitrary temperature
and magnetic field. The series coefficients will be presented
elsewhere and made available on the web.

To begin our comparison with the experimental data, we consider first the
uniform susceptibility in zero field. We note that
there are some deviations between the susceptibility measurements
of different groups. We will consider here the measurements of
Hammar {\sl et al.} \cite{CuHpCl2} and use them to fix exchange parameters 
within the three models. In Fig. 1 we compare the experimental 
data with the models with parameters $J_\perp=13K$, $J_l=3.5K$ for
the ladder, $J_\perp=13.5K$, $J_a=5.1K$ for the alternating chain
and $J_\perp=13K$, $J_2=1.9K$ for the bilayer. The fit was obtained with
$g=2.04$ as given by  Hammar {\sl et al.} \cite{CuHpCl2}. We note that the
agreement is excellent for all three models. 
Different models essentially reproduce
the experimental data much better than the difference in
different experimental results.

\begin{figure}
  \protect\centerline{\epsfig{file=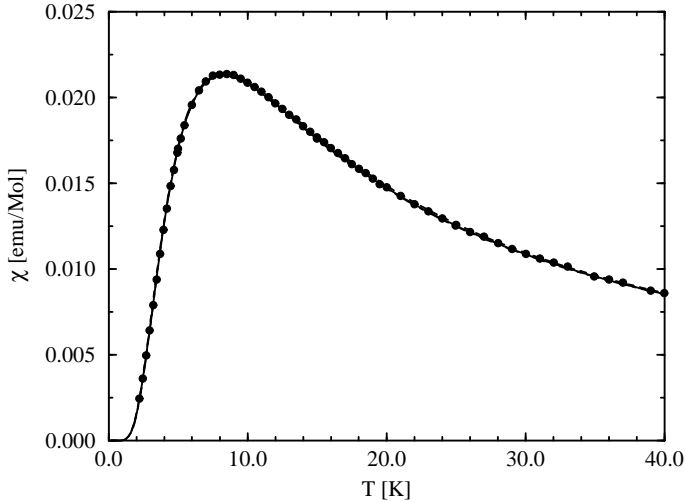, height=3in}}
  \protect\caption{Uniform susceptibility $\chi$ in zero field per spin 
           vs. temperature $T$. The lines are partial sums of series 
           for the three models discussed in the text. The curves are
           almost indistinguishable and provide an excellent fit to the
           experimental data of Hammar {\sl et al.} } 
\label{fig:chi_H=0}
\end{figure}

We note that similar ambiguity has
also been encountered in studying the system (VO$_2$)P$_2$O$_7$ \cite{VPO},
where many initial measurements were interpreted in terms of
a spin-ladder model but more recent neutron scattering measurements
suggest that the alternating chain model is more appropriate.

In Fig. 2 the zero-field specific heat measurements \cite{CuHpCl2}
are compared with the
theoretical models. We note that while the shape of the experimental spectra
is very similar and its activated low temperature behavior and peak position
are well reproduced by the series, the overall quantitative agreement with
theory is missing. This maybe due to incomplete background subtractions or the
presence of non-magnetic phases in the experimental samples.

\begin{figure}
  \protect\centerline{\epsfig{file=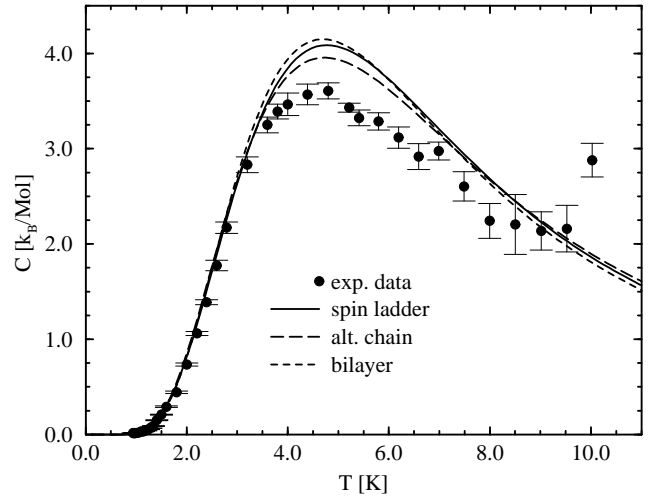, height=3in}}
  \protect\caption{Specific heat $C$ in zero field per spin 
           vs. temperature $T$. The lines are partial sums of series 
           for the three models discussed in the text compared with 
           experimental data of Hammar {\sl et al.} } 
\label{fig:C_H=0}
\end{figure}

Before we move on to field dependent measurements, we turn to the calculation
of the critical field.  The triplet dispersion has been calculated to
high orders by zero temperature dimer expansions
\cite{weihong,Ladder}.  For the parameters chosen, the gap is
determined to be $10.0\, K$ for the ladder model, $10.24\, K$ for the
alternating spin chain and $9.1\, K$ for the bilayer model.  Which
taking $g=2.04$ translates into critical fields of $7.30\, T$, $7.45\,
T$ and $6.65\, T$ respectively.  Given that experimental observation
of critical fields of 7.2(1) Tesla by Hammar {\sl et al.}
\cite{CuHpCl2} and approximately 7.7 Tesla by Chaboussant {\sl et al.}
\cite{CuHpCl1}, this argues in favor of the one-dimensiopnal models.

In Fig. 3 we compare the Knight shift measurements of Chaboussant {\sl
et al.} \cite{CuHpCl3} with the M/H ratio calculated for the ladder
model. Again the two models were found to be hardly distinguishable.
down to quite low temperatures. We keep the exchange constants from
the previous comparisons, but the vertical scale is arbitrary.  Given
that the zero field data of the two groups disagree somewhat, the
agreement, found here, is quite good. The convergence of these
expansions breaks down once the gap closes but only at fairly low
temperatures, when the correlation length becomes large.

\begin{figure}
  \protect\centerline{\epsfig{file=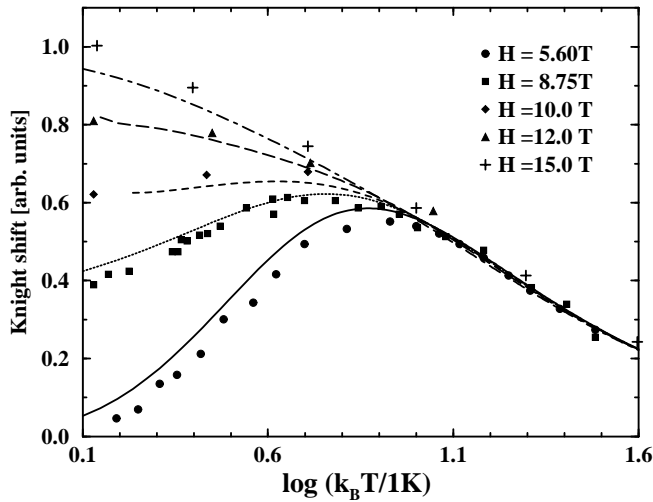, height=3in}}
  \protect\caption{Knight shift vs. $log(T)$. Partial sums of series
           for the ladder model at various magnetic fields compared 
           with experimental data of Chaboussant {\sl et al.} } 
\label{fig:Knight}
\end{figure}

In Fig. 4 we show the temperature dependence of the magnetization
for the two models, when the system is tuned to the quantum
critical point. The difference is quite apparent. The low
temperature power-law behavior is known \cite{Sachdev} to be $M\propto T^{1/2}$
in d=1 and $M\propto T$ in d=2.
Although the asymptotic low temperature behavior may set in
only at extremely low temperatures the difference between the
two curves is quite apparent. In 1D the magnetization approaches
zero with an infinite slope, whereas in 2D it appears to do
so with a finite or zero slope. By checking consistency with linear
or square-root behavior one can distinguish the two cases quite clearly 
at temperatures as high as 4K.
The material CuHpCl undergoes a
3-dimensional phase transition around 1K, hence, the window for
observing quantum critical behavior maybe very limited.
We note that in recent measurements of the nuclear relaxation
rates for this material Chaboussant {\sl et al.} observe a fairly
sharp upturn around 5K \cite{CuHpCl3}, which they interpret as evidence
for quantum critical behavior. This crossover temperature scale is
consistent with what we find here.

\begin{figure}
  \protect\centerline{\epsfig{file=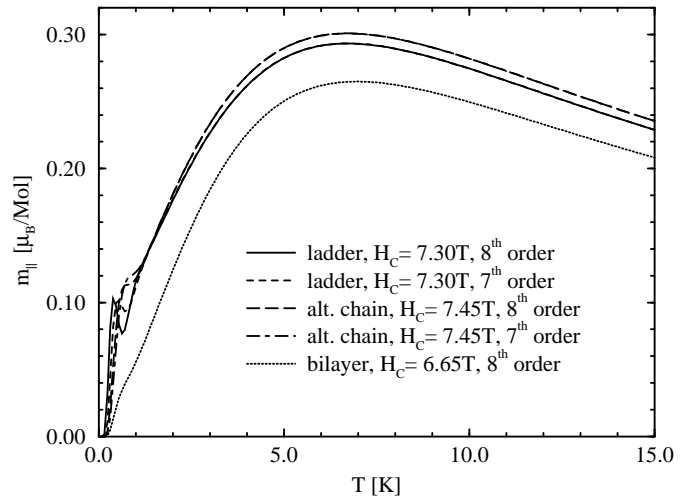, height=3in}}
  \protect\caption{Magnetisation $m_\parallel$ per spin 
           vs. temperature $T$ for the critical field $H_c$ 
           as discussed in the text.}
\label{fig:mz_Hc}
\end{figure}

Assuming the spin-ladder model, 
a mean-field treatment \cite{mean-field} of the transition temperature,
would lead to interchain coupling of the same order of magnitude
as the coupling along the polymeric chains (of order 1K).
This large interchain coupling would explain the absence of Van Hove
singularities in the powder neutron diffraction.
On the other hand, if the stronger couplings make it a quasi-2D system, 
with let us say weaker coupling along the polymeric chains, there
could be a finite temperature Kosterlitz-Thouless phase
transition, which could
turn into a 3D phase transition even with very weak 3D couplings.

In conclusion, in this paper we have presented finite temperature strong
coupling expansions for the uniform susceptibility,
magnetization, internal energy and specific
heat of a number of dimerized spin models at arbitrary 
temperatures and fields.
These calculations should be quite useful in 
experimental determination of exchange parameters for 
a class of magnetic materials.
We also showed that uniform thermodynamic measurements in zero field
are not sufficient for determining the nature of the spin Hamiltonian
and the dimensionality of the system. We found that the susceptibility
data for the material CuHpCl can be equally well fit by a
number of different models. 

We have also shown that field tuning the system to the
critical point, may provide a clear way to determine
the effective dimensionality of the spin system and to
study the asociated quantum critical phenomena. Direct measurements of
the spin dispersion by neutron diffraction on single crystals
should shed more light on the nature of the coupling constants
in this material.

The series can be obtained on the WWW. The access is via 
http://brahms.physik.uni-bonn.de/\~\ norbert/series/series.html

Acknowledgements: 
We would like to thank Daniel Reich for making the
experimental data available to us and to G. Chaboussant
for sending us a copy of his manuscript prior to publication.
One of us (NE) acknowledges the hospitality of 
the University of California at Davis where the early stages of this
work were done. 
This work is supported in part by
the US National Science Foundation under Grants No.
DMR-96-16574.

\pagebreak

\end{document}